\documentclass[osajnl,twocolumn,showpacs,10pt]{revtex4-1} 
\usepackage{amsmath,amssymb,graphicx}
\usepackage[colorlinks=true,urlcolor=blue,citecolor=blue,linkcolor=blue]{hyperref}
\usepackage{color}
\definecolor{blaa}{RGB}{153,153,255}
\definecolor{filtered}{RGB}{153,0,51}

\usepackage{amsmath}
\usepackage{amssymb}
\usepackage{epstopdf}
\usepackage[T1]{fontenc}
\usepackage{times}

\begin{document}

\title{Steerable optical tweezers for ultracold atom studies}

\author{K. O. Roberts}
\author{T. McKellar}
\author{J. Fekete}
\affiliation{Jack Dodd Centre for Quantum Technology,
Department of Physics, University of Otago, Dunedin, New Zealand}
\author{A. Rakonjac}
\affiliation{Jack Dodd Centre for Quantum Technology,
Department of Physics, University of Otago, Dunedin, New Zealand}
\author{A. B. Deb}
\author{N. Kj{\ae}rgaard}
\email{Corresponding author: niels.kjaergaard@otago.ac.nz}
\affiliation{Jack Dodd Centre for Quantum Technology,
Department of Physics, University of Otago, Dunedin, New Zealand}

\begin{abstract}We report on the implementation of an optical tweezer system for controlled transport of ultracold atoms along a narrow, static confinement channel. The tweezer system is based on high-efficiency acousto-optical deflectors and offers two-dimensional control over beam position. This opens up the possibility for tracking the transport channel when shuttling atomic clouds along the guide, forestalling atom spilling. Multiple clouds can be tracked independently by time-shared tweezer beams addressing individual sites in the channel. The deflectors are controlled using a multichannel direct digital synthesizer, which receives instructions on a sub-microsecond time scale from a field-programmable gate array. Using the tweezer system, we demonstrate sequential binary splitting of an ultracold $\rm^{87}Rb$ cloud into $2^5$ clouds.
\end{abstract}


\maketitle 
\begin{figure*}[t!]
\begin{center}
    \includegraphics[width=0.75\textwidth]{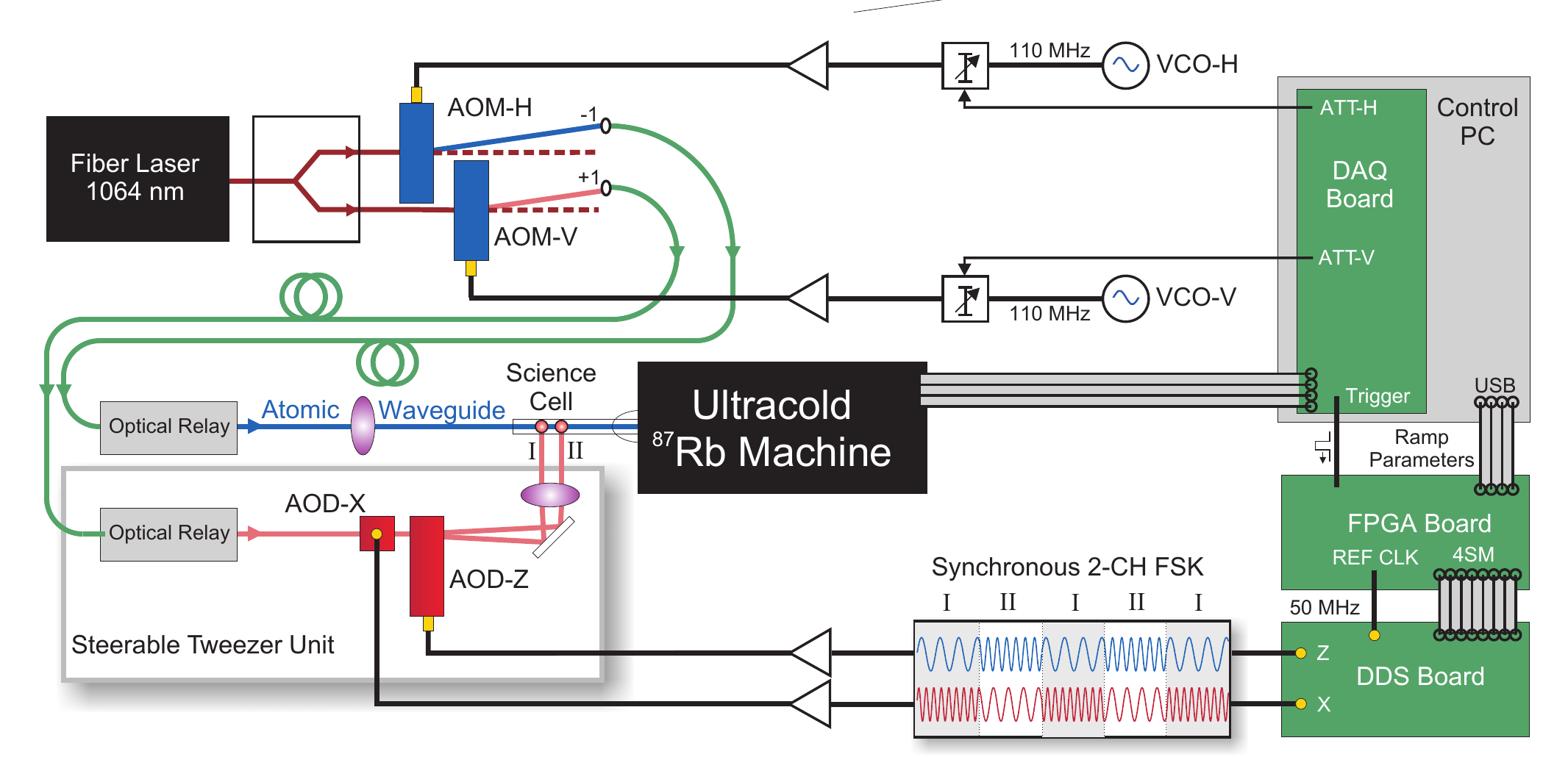}
	\caption{Simplified schematic of experimental setup showing essential functionalities.\label{fig:setup}}
	\label{fig:setup}
\end{center}
\end{figure*}

Many applications in cold atom quantum technology require control over the positions and momenta of atomic clouds or single atoms.
The ability to precisely move clouds of atoms around in space may, for example, unlock their use as probes for sampling surfaces and microscopic structures \cite{Gierling2011} or for mapping out magnetic fields \cite{Fatemi2010}. Several protocols for quantum information processing rely on deterministic rearrangement and controlled transport of neutral atoms \cite{Muldoon2012}, where, e.g., quantum logic gates can be formed through the precise movement of atomic qubits \cite{Beugnon2007}. Transport of atoms also emerges as a crucial step in the everyday operation of contemporary cold atom experiments for the purpose of shuttling atoms to a region of higher vacuum or increased optical access \cite{Greiner2001, Gustavson2001,Lewandowski2003,Schmid2006}.

Techniques for confining and manipulating polarizable particles by means of optical dipole forces from far-detuned laser beams have gained interdisciplinary importance after the seminal work of Ashkin, with applications ranging from folding of DNA to single atom trapping \cite{Ashkin2000}. A focused red-detuned laser beam acts as an optical tweezer, trapping atoms at the waist \cite{Chu1986}. By actively shifting the laser beam focus axially \cite{Gustavson2001} or by steering the laser beam transversely \cite{Beugnon2007,Shin2004}, the positions of atoms may be controlled; we note that steering can also be accomplished using blue-detuned light \cite{Zawadzki2010,Fatemi2010}.
Steerable tweezers have been widely applied, particularly within  the field of micro- and biological particles. The figures-of-merit of various approaches are, for example, reviewed
in \cite{Neuman2004}. Devices for tweezer steering include galvo and piezo-deflected mirrors, electro-optic deflectors, acousto-optic deflectors (AODs), as well as spatial light modulators. In the context of 	micro- and biological particles, manipulation of \textit{multiple} optical traps was first demonstrated two decades ago using time-averaged confinement from a rapidly galvo scanned laser beam \cite{Visscher1993,Sasaki1991}. Work using AODs \cite{Molloy1998} for time-averaged confinement quickly followed and was successfully extended to micromanipulation of ultracold atoms in multiplexed discrete traps (up to three) \cite{Onofrio} and for ``painting'' arbitrary potentials \cite{Henderson2009}.

In this Letter, we demonstrate micro- and macro-manipulation of ultracold atoms in a horizontal static dipole trap waveguide, crossed by vertical steerable optical tweezers.
The tweezers are steered using a time-shared AOD approach, with deflections controlled by the radpidly alternating frequency output from programmable direct digital synthesizers (DDSs) operated in a multiple frequency-shift keying (MFSK) mode. This platform offers a large dynamic range for position control. We report on the \textit{uniform} partition of a microkelvin cold cloud into 32 daughter clouds (requiring microscopic control), occupying a 4~mm long array (requiring macroscopic control) of equidistant sites. Using dual-axis control over the steerable tweezers, we also demonstrate the capability of tracking the static waveguide. With tracking engaged, an increased trap lifetime is observed towards the axial extremum positions covered in the channel.

\begin{figure}[b!]
\begin{center}
    \includegraphics[width=0.9\columnwidth]{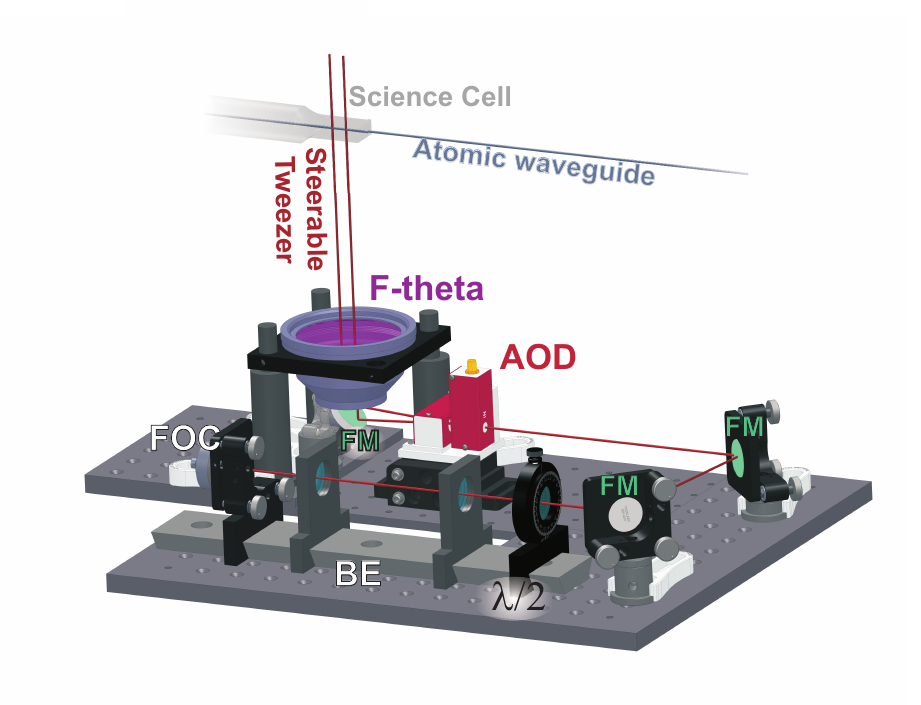}
	\caption{Steerable optical tweezer unit. \label{fig:setupsolidworks}}
\end{center}
\end{figure}
A schematic of our experimental setup is shown in Fig.~\ref{fig:setup}.
Light for a red-detuned optical tweezer potential is derived from a 1064~nm fiber laser (IPG Photonics, YLR-50-1064-LP-SF).
The laser has a single frequency output ($\sim100\,$kHz linewidth) and is typically operated at $12\,$W optical power, which we
divide into two beams by an arrangement of half-wave retarders and polarizing beam splitters. The power of each beam (which we shall denote by H and V respectively) is regulated by acousto-optic modulators (AOMs). Both AOMs are driven at a fixed frequency of 110~MHz by amplified voltage controlled oscillators (VCOs) at an rf power level determined by voltage variable attenuators at the VCO outputs. Our main experiment control PC sets the optical power diffracted by the AOMs via these attenuators by means of a 16-bit data acquisition board (DAQ; NI PCI-6733). Alignments of the H-AOM and V-AOM are optimized for diffraction into $-1$st and +1st order, respectively, and the linearly polarized diffracted beams, coupled into polarization-maintaining single mode fibers. The output beam of the H-branch enters our science cell in the horizontal plane along the axial direction ($z$-axis) of a Ioffe Prichard (IP) style atom trap and is focused to a waist of 60~$\rm \mu m$ at the trap center by means of an f=250~mm lens. This beam constitutes the horizontal atomic waveguide along which we want to move atoms. The output of the V-branch enters our steerable optical tweezer unit, integrated onto a breadboard (see Fig.~\ref{fig:setupsolidworks}). This optomechatronic unit consists of: an optical fiber output collimator (FOC) followed by a Keplerian telescope for (optional) beam expansion (BE), a half wave retarder ($\lambda/2$) for polarization control, folding mirrors (FMs), a two-axis acousto-optic deflector (AOD) mounted on a four-axis tilt aligner (New Focus 9071), and a f=130~mm  F-theta lens (Eksma Optics).

The two-axis AOD (AA Sa, DTSXY-250-1064) is characterized by a nominal angular beam deflection range of $2.8$ degrees along two directions, denoted by $X$ and $Z$ respectively.  For our application, we aim at covering a large range in the $Z$ dimension while varying the frequency corresponding to $X$ by relatively small amounts. We measured optical the efficiency of our AOD to exceed $60\,\%$ over a $Z$-frequency range of $f_Z$=65-85~MHz, for $X$-frequencies within the interval $f_X$=70-76~MHz. The tweezer unit is typically operated at an optical power corresponding to 2.8~W after the FOC. For a collimated input beam diameter of $\sim2$~mm, the access time of the acoustic wave through the AOD aperture was measured to be $2\,\mu$s, imposing an upper bound of $500$~kHz for the rate at which we can perform FSK.
\begin{figure}[t!]
\begin{center}
    \includegraphics[width=\columnwidth]{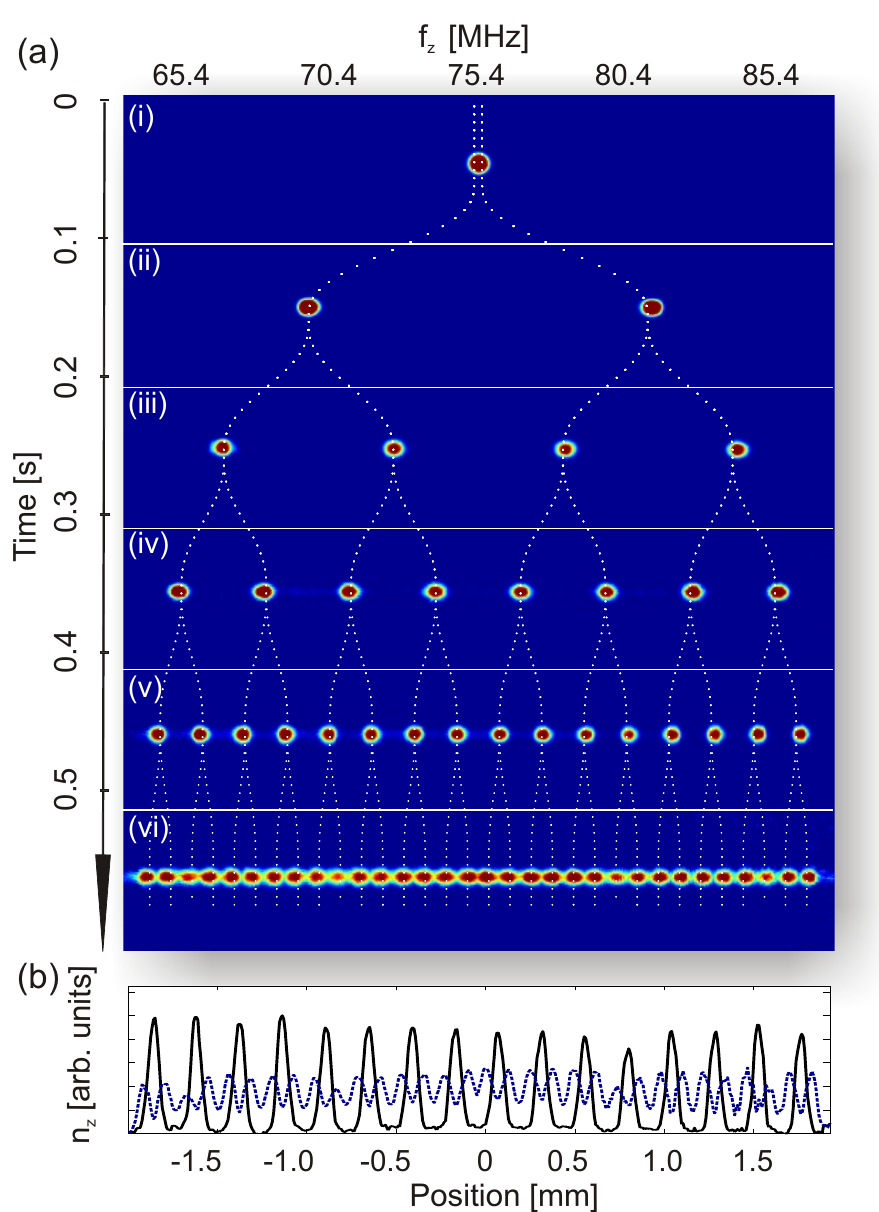}
	\caption{(a) (\href{http://www.physics.otago.ac.nz/staff_files/nk/files/sites32.mov}{Media 1}) Absorption images showing the production (ii-vi) of 2,4,8,16, and 32 separate atomic samples in a linear array from a single cloud in a double-well (i) . The white bifurcating curve overlaying the frames shows $f_Z$ versus time. (b) Axial atom number
distributions for arrays with 16 (solid black line) and 32 sites (dashed blue line).}
	\label{fig:magnetic}
\end{center}
\end{figure}

The output beam from the steerable tweezer unit enters the science cell along a vertical direction, intersecting the horizontal waveguide at the IP trap center position for an AOD frequency pair $(f_X,f_Z)=(72.4,75.4)$~MHz. The F-theta lens at the tweezer unit output focuses the V-beam to a waist of 40~$\rm\mu m$ at the intersection point. Over the range accessible via the AOD, this lens (optimized for flat field imaging) provides a close to constant waist in the horizontal plane defined by the H-beam guide. For the AOD-to-lens separation in our implementation the beam propagates paraxially after the F-theta lens and the accessible area corresponds to $6\times 6$~mm$^2$.

As a first example showcasing the flexibility of our system, we demonstrate the splitting and transport
of multiple atomic clouds along the horizontal waveguide. A commercially available programmable frequency source (WieserLab FlexDDS) controls the V-beam position via $f_Z$. An ultracold sample of $\rm^{87}Rb$ prepared in the IP trap \cite{Sawyer2012} is loaded into a crossed dipole trap set up by the H-and V-beams. Before commencing any movement, the typical atom number is $3\times10^6$ and the temperature of the sample is below $1\,\mu$K. Initially, the V-beam forms a time-averaged double-well potential along the $z$-axis as a result of performing FSK of $f_Z$ at a rate of 75~kHz between two closely spaced frequencies about $f_Z=75.4$~MHz. By concurrently ramping the high and low frequency of the FSK up and down, respectively, the sample is split into two. We employ a frequency ramp that makes the beam foci trace out a minimum jerk cost trajectory \cite{FLASH1985}, causing no discernable heating or atom loss. Figure~\ref{fig:magnetic}(a) shows arrays of equidistantly spaced clouds produced along the horizontal waveguide by replicating this splitting procedure up to four additional times for daughter clouds. The approach using a DDS frequency source allows for a truly macroscopic intercloud separation while retaining control at the microscopic level such that clouds can be split evenly and consistently as evidenced in Fig.~\ref{fig:magnetic}(b).

\begin{figure}[tb!]
\begin{center}
    \includegraphics[width=0.75\columnwidth]{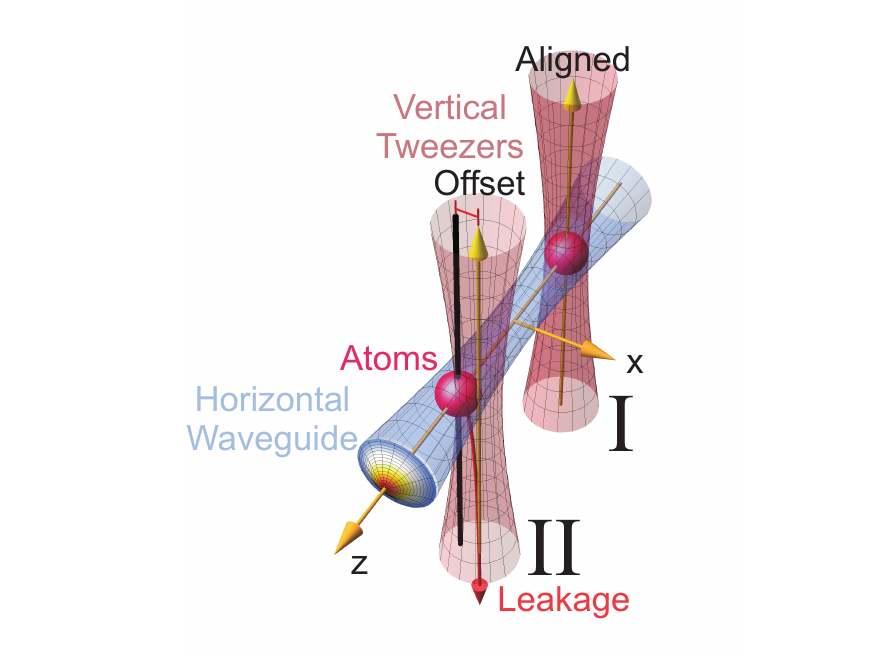}
	\caption{Crossed optical dipole traps and trapping geometry. Example showing vertical tweezer I aligned to intersect the axis of the horizontal waveguide, while tweezer II is offset in the $x$-direction. The latter leads to leakage and loss of atoms. }
	\label{fig:tweezermathematica}
\end{center}

\end{figure}

The trapping potential for multiple clouds is created by time-sharing the V-beam between discrete sites via FSK of $f_Z$. For macroscopically
spaced sites along the microscopically wide H-beam guide, it is generally a non-trivial task to ensure perfect alignment where the optical axes of the H-and V-beams intersect at every site. If there is an offset between the axes on the scale of a beam waist ($\sim50\mu$m), this may lead to leaking of atoms out of a crossed trap as illustrated in Fig.~\ref{fig:tweezermathematica}. As a solution, we demonstrate how the two-axis AOD in conjunction with a synchronous two-channel FSK drive can be used to track the waveguide.

For two-channel FSK, we employ a multichannel DDS (Analog Devices eval-AD9959) programmed by a field-programmable gate array (FPGA; Xilinx Spartan-3AN).
The reference oscillator of the DDS is provided by the FPGA internal clock ($50\,$MHz), multiplied up to $500\,$MHz by an on-board phase-locked loop.
Communication from the FPGA to the DDS is run in 4-bit Serial Mode (4SM) \cite{9959}, clocked at $50\,$MHz.
The frequency output of each channel  is defined on a 0-250~MHz interval by a 32-bit frequency tuning word (FTW) . Using 4SM, dual channel FSK can be performed at a rate of $\sim1.5$~MHz.
Frequency trajectories for each DDS channel are created by calculating new FTWs in real time on the FPGA based on parameters pre-loaded in an on-board memory via USB.
The algorithms to read the ramp parameter data from memory, calculate the frequency ramps, and communicate to the DDS are all implemented in the FPGA design, while excution of the FPGA administered frequency sweep is triggered by the main experimental control PC.

\begin{figure}[tb!]
\begin{center}
    \includegraphics[width=0.9\columnwidth]{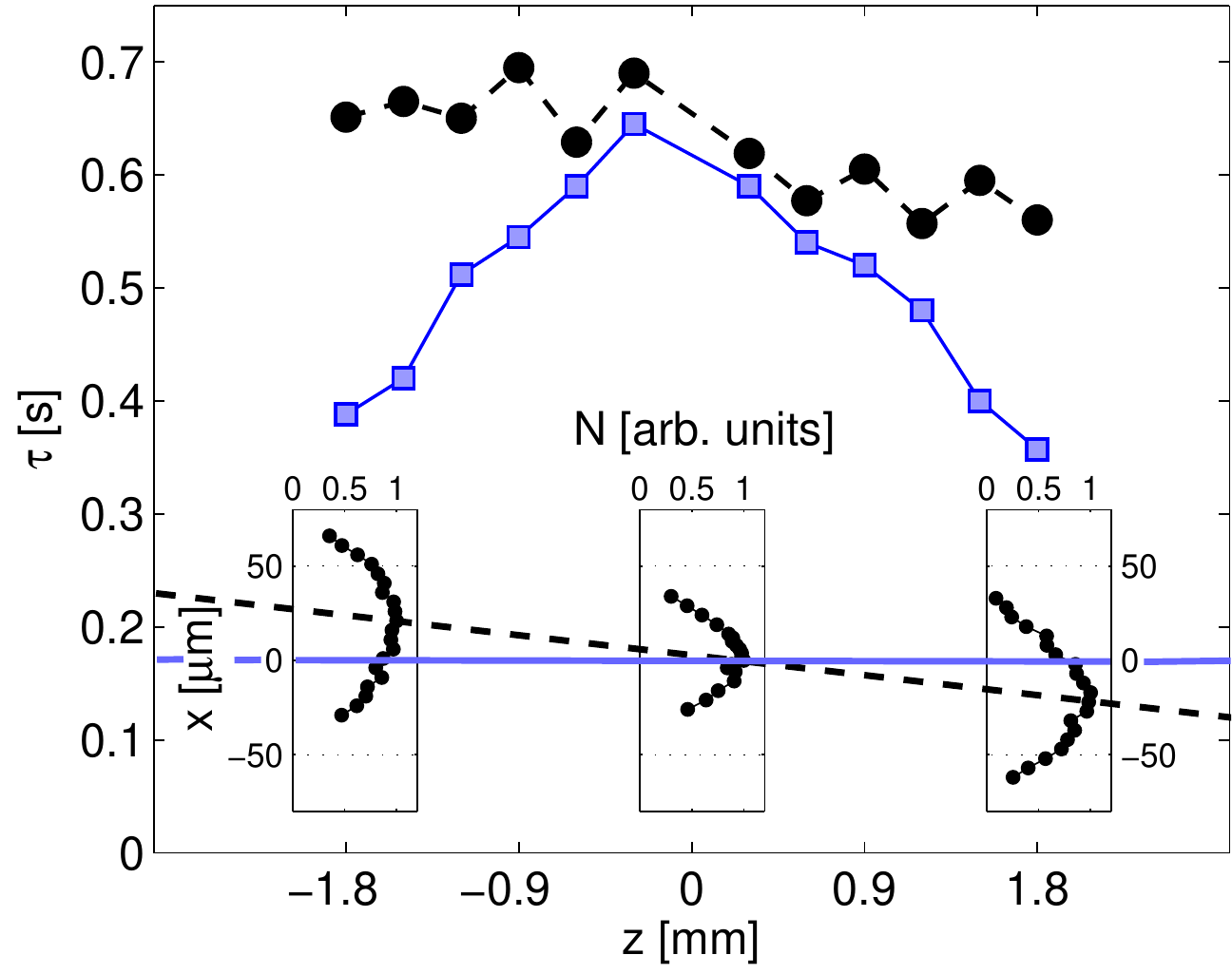}
	\caption{Crossed-trap lifetimes along the waveguide; {\scriptsize\textcolor{blaa}{$\blacksquare$}}, single channel FSK with  $f_X=72.4$~MHz (fixed) optimized for $z=0$; {\large$\bullet$}, dual-channel FSK with $f_X$ varying linearly with $z$ and optimized lifetimes at $z=\pm1.8$~mm. Optimizations were based on the data displayed in the inset: number of atoms left $100$~ms after loading site versus $f_X$ at $z =\{ -1.8, 0, 1.8\}$~mm. Optimized $f_X(z)$ trajectories for single and dual-channel FSK shown as solid blue and dashed black lines, respectively.\label{fig:xtoggle}}
\end{center}
\end{figure}

Fig~\ref{fig:xtoggle} shows the $1/e$ lifetime $\tau$ for crossed trap sites along the horizontal waveguide for different final separations, measured after performing a bipartition of a cloud [see Fig.~\ref{fig:magnetic}(a,ii)] . When keeping $f_X$ fixed to the value that optimizes $\tau$ at $z=0$, we observe a clear decrease in lifetime with increasing $z$-distance to the origin. The loss of particles can be greatly mitigated using synchronous dual-channel FSK for tracking the waveguide as described above. By using a linear (in $z$) tracking function for $f_X$ that optimizes the lifetimes at $z=\pm 1.8$~mm (our widest separation) we observe an increased lifetime for all in-between points. The $x$-corrections at the points at these extreme positions correspond to only $\sim\pm20~\mu$m, indicating a fortuitously good initial passive alignment. Because of this, the data presented in Fig.~\ref{fig:xtoggle} were acquired with a relatively low power (0.45~W) in the H-beam to make the effect evident; for higher H-beam powers ({$\gtrsim1$~W), our passive alignment is in fact sufficiently good that the V-Beams will not pull an appreciable number of atoms out of the waveguide for the splitting distances considered here. The value of the tracking method should be clear, however, and even for our close passive alignment (within $\sim10$~mrad), it would take less than triple our maximum splitting for the V-beams and H-beam to lose their spatial overlap.

For multiplexed arrays with $>16$ sites, a quality improvement was also observed, when engaging tracking of $f_X$, even for high H-beam powers (because of this, all of the frames of Fig.~\ref{fig:magnetic} and \href{http://www.physics.otago.ac.nz/staff_files/nk/files/sites32.mov}{Media 1} were aquired using $f_X$-tracking for overall consistency). In the untracked case, atoms are not lost by funneling down a V-beam, but rather along the H-beam guide due to lowered effective time-shared power at a site.

In previous work \cite{Rakonjac2012}, we demonstrated an optical collider using a \textit{single} AOD with a \textit{continuous} two-frequency drive. Here, the spilling of atoms along a vertical tweezer beam when separating clouds was identified as one of the main performance limitations.
Several other AOD implementations \cite{Shin2004,Zawadzki2010,Trypogeorgos2013} of multiple traps and designed, arbitrary potentials for atoms have made use of static distributions of trapping light contrasting the ``toggled'' time-sharing method adopted in the present Letter.  We believe the toggled approach has distinct advantages for trapping geometries along a waveguide: in a continuous drive approach, e.g., tracking of sites would require half of the optical power to be deposited in ``ghost'' traps outside the waveguide.

In conclusion, we have demonstrated that inexpensive multi-channel DDS boards, which have become available in recent years, provide an efficient means for implementing multiple time-averaged traps and their micro and macro-manipulation. Such devices offer a frequency resolution of a fraction of a Hertz as well as full control over the phase of waveforms; in the present work, e.g., phase continuous FSK was employed. Specifically, we have reported on a two-axis steerable optical tweezer unit with a DDS FSK drive capable of tracking an atomic waveguide. The tweezers were used to realize regular arrays of ultracold atoms (up to 32 sites) covering a distance of almost half a centimeter. We are currently investigating the prospects of employing our system for controlled, site-selective evaporative cooling to quantum degeneracy and for schemes in the context of magnetic gradiometry \cite{Deb2013}.

This work was supported by The Marsden Fund (Contract No. UOO1121)
and a University of Otago Research Grant.

\end{document}